\documentclass[runningheads]{llncs}
\usepackage[T1]{fontenc}
\usepackage{graphicx}
\usepackage{multirow}
\usepackage{multicol}
\usepackage{tabularx}
\usepackage{subcaption}
\usepackage{makecell}
\usepackage{tabularray}
\usepackage{float}
\usepackage{tikz}
\usepackage[title]{appendix}
\usetikzlibrary{shapes,decorations,arrows,calc,arrows.meta,fit,positioning}
\tikzset{
    -Latex,auto,node distance = .2 cm and .2 cm, semithick,
    state/.style ={ellipse, draw, minimum width = 0.5cm},
    point/.style = {circle, draw, inner sep=0.01cm,fill,node contents={}},
    bidirected/.style={Latex-Latex,dashed},
    el/.style = {inner sep=1pt, align=center, sloped}
}
\newcommand*{\myfont}{\fontfamily{lmtt}\selectfont}
\DeclareTextFontCommand{\textmyfont}{\myfont}

\begin{document}

\title{Counterfactual Fairness Evaluation of Machine Learning Models on Educational Datasets}
\titlerunning{Counterfactual Fairness Evaluation on Educational Datasets}

\author{Woojin Kim\inst{1} \and Hyeoncheol Kim\inst{1*}}

\authorrunning{W Kim.}

\institute{Department of Computer Science and Engineering, Korea University, South Korea \\
\email{\{woojinkim1021, harrykim\}@korea.ac.kr}}

\maketitle              

\begin{abstract}
As machine learning models are increasingly used in educational settings, from detecting at-risk students to predicting student performance, algorithmic bias and its potential impacts on students raise critical concerns about algorithmic fairness. Although group fairness is widely explored in education, works on individual fairness in a causal context are understudied, especially on counterfactual fairness. This paper explores the notion of counterfactual fairness for educational data by conducting counterfactual fairness analysis of machine learning models on benchmark educational datasets. We demonstrate that counterfactual fairness provides meaningful insight into the causality of sensitive attributes and causal-based individual fairness in education.

\end{abstract}

\section{Introduction}
Machine learning models are increasingly implemented in educational settings to support automated decision-making processes. Such applications ranges from academic success prediction\cite{pallathadka2023classification,yaugci2022educational}, at-risk detection\cite{lakkaraju2015machine}, automated grading\cite{taghipour2016neural}, knowledge tracing\cite{piech2015deep} and personalized recommendation\cite{zhou2018personalized}. However, the application of machine learning models to automate decision-making in high-stakes scenarios calls for consideration of algorithmic bias\cite{baker2022algorithmic}. In education, predictive models have been shown to exhibit lower performance for students from underrepresented demographic groups \cite{sha2022leveraging,bird2024algorithms,cock2023protected,jiang2021towards,pan2024examining}.

The majority of research on fairness in education focuses on group fairness\cite{sha2022leveraging,jiang2021towards}, while works on individual fairness are limited to aiming for similar treatment of similar individuals\cite{hu2020towards,deho2022existing}. Under context where students' demographics causally shape their education \cite{Webbink2005Causal,delaney2021gender,Li2024survey}, taking causality in consideration of fairness is crucial. Causal fairness asserts that it is unfair to produce different decisions for individuals caused by factors beyond their control\cite{loftus2018causal}. In this sense, algorithmic decisions that impact students should eliminate the causal effects of uncontrollable variables, such as race, gender, and disability.

Group and individual fairness definitions have certain limitations, and the inherent incompatibility between group and individual fairness presents challenges\cite{binns2020apparent,makhlouf2021machine,zhou2022group,xu2024compatibility}. Group fairness can mask heterogeneous outcomes of individuals by using group-wise averaging measurements\cite{binns2020apparent,makhlouf2021machine}. While group fairness may be achieved, it does not ensure fairness for each individual\cite{zhou2022group}. Furthermore, ignoring individual fairness in favor of group fairness can result in algorithms making different decisions for identical individuals\cite{long2023individual}. Individual fairness faces difficulty in selecting distance metrics for measuring the similarity of individuals and is easily affected by outlier samples\cite{xu2024addressing}. 

Based on the limitations of group and individual fairness notions, we empirically investigate the potential of counterfactual fairness on educational datasets. Counterfactual fairness ensures that the algorithm's decision would have remained the same when the individual belongs to a different demographic group, other things being equal\cite{kusner2017counterfactual}. Counterfactual fairness promotes individual-level fairness by removing the causal influence of sensitive attributes on the algorithm's decisions. To the best of our current knowledge, the notion of counterfactual fairness has not been investigated in the educational domain. 

In this paper, we aim to answer the following research questions(RQ): 
\begin{enumerate}
  \item What causal relationships do sensitive attributes have in educational data?
  \item Does counterfactual fairness in educational data lead to identical outcomes for individual students regardless of demographic group membership?
  \item Does counterfactually fair machine learning models result in a performance trade-off in educational data?
\end{enumerate}

These questions are investigated by estimating a causal model and implementing a counterfactual fairness approach on real-world educational datasets. Section \ref{section2} introduces counterfactual fairness and algorithmic fairness in education. In Section \ref{section3}, we provide methodologies for creating causal models and counterfactual fairness evaluation metrics. We present the experiment result in Section \ref{section4}. In Section \ref{section5}, we discuss the key findings of our study, exploring their implications for fairness in educational data before concluding in Section \ref{section6}.

\section{Background}
\label{section2}

\subsection{Causal Model and Counterfactuals}

Counterfactual fairness adopts the Structural Causal Model(SCM) framework\cite{pearl2009causality} for the calculation of counterfactual samples. SCM is defined as a triplet $(U, V, F)$ where $U$ is a set of unobserved variables, $V$ is a set of observed variables, and $F$ is a set of structural equations describing how observable variables are determined. Given a SCM, counterfactual inference is to determine $P(Y_{Z \gets z}(U) | W=w)$, which indicates the probability of $Y$ if $Z$ is set to $z$(i.e. counterfactuals), given that we observed $W=w$. Imagine a female student with a specific academic record. What would be the probability of her passing the course if her gender were male while keeping all other observed academic factors constant? Counterfactual inference on SCM allows us to calculate answers to counterfactual queries by abduction, action, and prediction inference steps detailed in \cite{pearl2009causality}. 

\subsection{Counterfactual Fairness}

We follow the definition of counterfactual fairness by Kusner et al.\cite{kusner2017counterfactual}. 

\begin{definition}[Counterfactual Fairness]
\label{counterfactual}
Predictor $\hat{Y}$ is counterfactually fair if under any context $X=x$ and $A=a$,
\[ P(\hat{Y}_{A\leftarrow a} (U)=y | X=x, A=a)=P(\hat{Y}_{A\leftarrow a'}(U)=y | X=x, A=a), \]
for all y and for any value a' attainable by A.
\end{definition}

The definition states that changing $A$ should not change the distribution of the predicted outcome $\hat{Y}$. An algorithm is counterfactually fair towards an individual if an intervention in demographic group membership does not change the prediction. For instance, the predicted probability of a female student passing a course should remain the same as if the student had been a male. 

Implementing counterfactual fairness requires a causal model of the real world and the counterfactual inference of samples under the causal model. This process allows for isolating the causal influence of the sensitive attribute on the outcome.

Counterfactual fairness is explored in diverse domains, such as in clinical decision support \cite{wu2024fairness} and clinical risk prediction\cite{pfohl2019counterfactual,wastvedt2024intersectional} for healthcare, ranking algorithm\cite{piccininni2022counterfactual}, image classification\cite{dash2022evaluating,jung2025counterfactually} and text classification\cite{garg2019counterfactual}.

\subsection{Algorithmic Fairness in Education}

Most works on algorithmic fairness in education focus on group fairness\cite{sha2022leveraging,jiang2021towards}. The group fairness definition states that an algorithm is fair if its prediction performance is equal among subgroups, specifically requiring equivalent prediction ratios for favorable outcomes. Common definitions of group fairness are Equalized Odds\cite{hardt2016equality}, Demographic Parity\cite{dwork2012fairness} and Equal Opportunity\cite{hardt2016equality}.

Individual fairness requires individuals with similar characteristics to receive similar treatment. Research on individual fairness in education focuses on the similarity. Marras et al.\cite{marras2022equality} proposed a consistency metric for measuring the similarity of students' past interactions for individual fairness under a personalized recommendation setting. Hu and Rangwala\cite{hu2020towards} developed a model architecture for individual fairness in at-risk student prediction task. Doewes et al.\cite{doewes2022individual} proposed a methodology to evaluate individual fairness in automated essay scoring. Deho et al.\cite{deho2022existing} performed individual fairness evaluation of existing fairness mitigation methods in learning analytics. 

There have been attempts to understand causal factors influencing academic success. Ferreira de Carvalho et al.\cite{Ferreira2018Applying} identifies causal relationships between LMS logs and student's grades. Zhao et al.\cite{zhao2017estimating} propose Residual Counterfactual Networks to estimate the causal effect of an academic counterfactual intervention for personalized learning. To the best of our knowledge, the notion of algorithmic fairness under causal context, especially under counterfactual inference in the educational domain remains unexplored.

\newcolumntype{y}{>{\hsize=.33\hsize\centering\arraybackslash}X}
\newcolumntype{t}{>{\hsize=.16\hsize\centering\arraybackslash}X}
\renewcommand\tabularxcolumn[1]{m{#1}}
\newcolumntype{n}{>{\hsize=.13\hsize\centering\arraybackslash}X}
\newcolumntype{s}{>{\hsize=.20\hsize\centering\arraybackslash}X}
\newcolumntype{w}{>{\hsize=.62\hsize\raggedright\arraybackslash}X}

\begin{table}[t]
    \caption{Feature descriptions of Law School and OULAD datasets. Student Performance dataset descriptions are provided in Table \ref{dataset:student performance} of Appendix \ref{append_data}.}
    \label{datasets}
    \centering
    \begin{tabularx}{\textwidth}{n s s w}
        \hline
        Data & Feature & Type & Description\\
        \hline
        \multirow{5}{*}{Law} & \textmyfont{gender} & binary & the student's gender \\
                             & \textmyfont{race} & binary & the student's race \\
                             & \textmyfont{lsat} & numerical & the student's LSAT score \\
                             & \textmyfont{ugpa} & numerical & the student's undergraduate GPA \\
                             & \textmyfont{zfygpa} & numerical & the student's law school first year GPA \\
        \hline
        \multirow{9}{*}{OULAD} & \textmyfont{gender} & binary & the student's gender \\
                               & \textmyfont{disability} & binary & whether the student has declared a disability \\
                               & \textmyfont{education} & categorical & the student's highest education level \\
                               & \textmyfont{IMD} & categorical & the Index of Multiple Deprivation(IMD) of the student's residence \\
                               & \textmyfont{age} & categorical & band of the student's age\\
                               & \textmyfont{studied credits} & numerical & the student's total credit of enrolled modules \\
                               & \textmyfont{final result} & binary & the student's final result of the module\\
        \hline
    \end{tabularx}
\end{table}

\vspace{-1em}

\begin{table}[t]
    \caption{Summary of datasets used for the experiment.}
    \label{data_info}
    \centering
    \begin{tabularx}{\textwidth}{y t t s t}
        \hline
        Data & Task & Sensitive Attribute & Target & \# Instances \\ 
        \hline
        Law School & Regression & \textmyfont{race}, \textmyfont{gender} & \textmyfont{zfygpa} & 20,798 \\
        OULAD & Classification & \textmyfont{disability} & \textmyfont{final result} & 32,593 \\
        Student Performance(Mat) & Regression & \textmyfont{gender} & \textmyfont{G3} & 395 \\
        Student Performance(Por) & Regression & \textmyfont{gender} & \textmyfont{G3} & 649 \\
        \hline
    \end{tabularx}
\end{table}

\newpage

\section{Methodology}
\label{section3}

We provide detailed description of experiment methodology for evaluating counterfactual fairness of machine learning models in education.

\subsection{Educational Datasets}

We use publicly available benchmark educational datasets for fairness presented in \cite{le2022survey}, which introduces four educational benchmark datasets for algorithmic fairness. Datasets are Law School\footnote{github.com/mkusner/counterfactual-fairness}\cite{wightman1998lsac}, Open University Learning Analytics Dataset (OULAD)\footnote{https://archive.ics.uci.edu/dataset/349/open+university+learning+analytics+dataset}\cite{kuzilek2017open} and Student Performance in Mathematics and Portuguese language\footnote{https://archive.ics.uci.edu/dataset/320/student+performance}\cite{cortez2008using}. Refer to Table \ref{datasets} and Table \ref{dataset:student performance} for description of dataset features used in the experiment. The summary of tasks and selection of sensitive attributes are outlined in Table \ref{data_info}. 

\begin{figure}[t]
    \centering
     \begin{subfigure}[b]{0.24\textwidth}
         \centering
         \includegraphics[width=\textwidth]{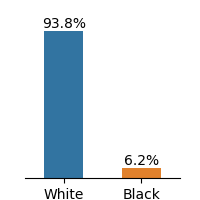}
         \caption{Law School}
         \label{fig:freq_law}
     \end{subfigure}
     \hfill
     \begin{subfigure}[b]{0.24\textwidth}
         \centering
         \includegraphics[width=\textwidth]{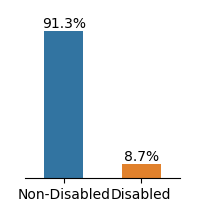}
         \caption{OULAD}
         \label{fig:freq_oulad}
     \end{subfigure}
     \hfill
     \begin{subfigure}[b]{0.24\textwidth}
         \centering
         \includegraphics[width=\textwidth]{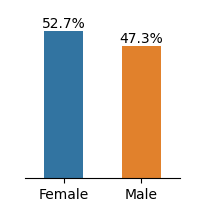}
         \caption{Mat}
         \label{fig:freq_mat}
     \end{subfigure}
     \hfill
     \begin{subfigure}[b]{0.24\textwidth}
         \centering
         \includegraphics[width=\textwidth]{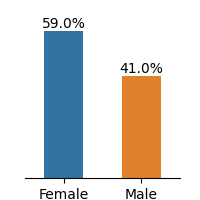}
         \caption{Por}
         \label{fig:freq_por}
     \end{subfigure}
    \caption{Frequency distributions of sensitive attributes in educational datasets.}
    \label{freq}
\end{figure}

The Law School dataset contains admission records of students at 163 U.S. law schools\cite{wightman1998lsac}. The dataset has demographic information of 20,798 students on race, gender, LSAT scores, and undergraduate GPA. We select \textmyfont{gender} and \textmyfont{race} as sensitive attributes and first-year GPA as the target for the regression task. 

The OULAD dataset, originating from a 2013-2014 Open University study in England, compiles student data and their interactions within a virtual learning environment across seven courses. We select \textmyfont{disability} as the sensitive attribute and \textmyfont{final result} as the classification target. The \textmyfont{gender} is not considered as our sensitive attribute because the preceding study\cite{hasan2022understanding} revealed that \textmyfont{gender} attribute does not have a causal relationship to student's \textmyfont{final result}. For this work, we only considered the module \textmyfont{BBB}(Social Science). 

The Student Performance dataset describes students' achievements in Mathematics and Portuguese language subjects in two Portuguese secondary schools during 2005-2006. The dataset provides details about students' demographics, and family backgrounds such as parent's jobs and education level, study habits, extracurricular activities, and lifestyle. We select \textmyfont{gender} as the sensitive attribute and \textmyfont{G3} as the target for the regression problem. Feature description of the dataset is presented in Appendix \ref{append_data}.

The dataset demonstrates imbalance between subgroups of sensitive attributes, presented in Fig. \ref{freq}. Law school and OULAD datasets exhibit an extreme imbalance in the selected sensitive attributes, while the gender attribute in Student Performance is less imbalanced.

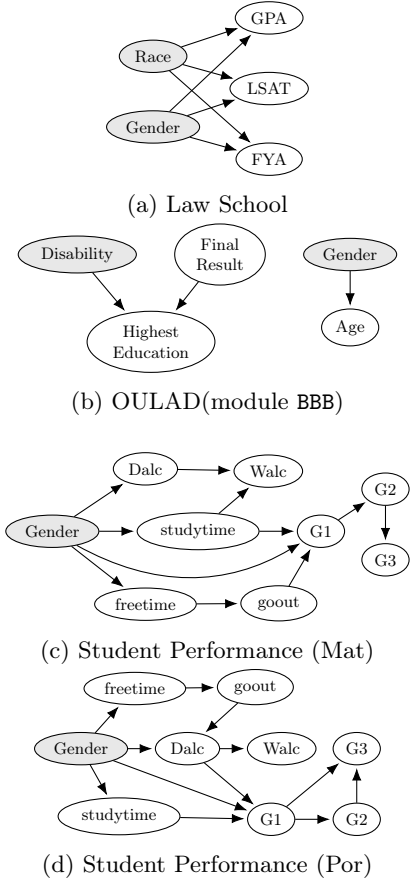
\begin{figure}
    \centering
        \begin{subfigure}[b]{0.45\textwidth}
            \centering
            \begin{tikzpicture}[> = stealth,  shorten > = 1pt,   auto,   node distance = 0.5cm, every node/.style={scale=.7}]
                \centering
                \label{graph:scm_law}
                \node[state,fill={rgb:black,1;white,10}] (race) {Race};
                \node[state,fill={rgb:black,1;white,10}] (gender) [below = of race] {Gender};
                \node[state] (gpa) [above right= 0.2 cm and 0.9 cm of race] {GPA};
                \node[state] (lsat) [below = of gpa] {LSAT};
                \node[state] (fya) [below = of lsat] {FYA};                
                \path (race) edge (gpa);
                \path (race) edge (lsat);
                \path (race) edge (fya);
                \path (gender) edge (gpa);
                \path (gender) edge (lsat);
                \path (gender) edge (fya);

            \end{tikzpicture}
            \caption{Law School}\label{dag_law}
        \end{subfigure}

        \begin{subfigure}[b]{0.45\textwidth}
            \centering
            \begin{tikzpicture}[> = stealth,  shorten > = 1pt,   auto,   node distance = 0.5cm, every node/.style={scale=.7}]
                \centering
                \label{graph:scm_oulad}
                \node[state,fill={rgb:black,1;white,10}] (disability) {Disability};
                \node[state][text width=1.5cm,align=center] (education) [below= of disability, xshift=1.4cm] {Highest Education}; 
                \node[state][text width=1.cm,align=center] (result) [right = of disability] {Final Result};
                \node[state,fill={rgb:black,1;white,10}] (gender) [right = of result] {Gender};
                \node[state] (age) [below = of gender] {Age};

                \path (disability) edge (education);
                \path (result) edge (education);
                \path (gender) edge (age);
            \end{tikzpicture}
            \caption{OULAD(module \textmyfont{BBB})}\label{dag_oulad}
        \end{subfigure}
        \par\bigskip
        \begin{subfigure}[b]{0.45\textwidth}
            \centering
            \begin{tikzpicture}[> = stealth,  shorten > = 1pt,   auto,   node distance = 0.5cm, every node/.style={scale=.7}]
                \centering
                \label{graph:scm_mat}
                \node[state,fill={rgb:black,1;white,10}] (gender) {Gender};
                \node[state] (studytime) [right = of gender] {studytime}; 
                \node[state] (freetime) [below= of studytime,xshift=-1cm] {freetime};
                \node[state] (dalc) [above= of studytime,xshift=-1cm,yshift=-0.2cm] {Dalc};
                \node[state] (g1) [right= of studytime] {G1};
                \node[state] (walc) [above= of g1,xshift=-1cm,yshift=-0.2cm] {Walc};
                \node[state] (goout) [below= of g1,xshift=-0.8cm] {goout};
                \node[state] (g2) [right= of g1,xshift=-0.4cm,yshift=0.8cm] {G2};
                \node[state] (g3) [below= of g2] {G3};

                \path (gender) edge (dalc);
                \path (gender) edge (freetime);
                \path (gender) edge (studytime);
                \path (gender) edge[bend right=30] (g1);
                \path (dalc) edge (walc);
                \path (freetime) edge (goout);
                \path (studytime) edge (walc);
                \path (studytime) edge (g1);
                \path (goout) edge (g1);
                \path (g1) edge (g2);
                \path (g2) edge (g3);
            \end{tikzpicture}
            \caption{Student Performance (Mat)}\label{dag_mat}
        \end{subfigure}

        \begin{subfigure}[b]{0.45\textwidth}
            \centering
            \begin{tikzpicture}[> = stealth,  shorten > = 1pt,   auto,   node distance = 0.5cm, every node/.style={scale=.7}]
                \centering
                \label{graph:scm_por}
                \node[state,fill={rgb:black,1;white,10}] (gender) {Gender};
                \node[state] (dalc) [right= of gender,xshift=-.2cm] {Dalc};
                \node[state] (freetime) [above= of dalc,xshift=-1cm,yshift=-0.2cm] {freetime};
                \node[state] (studytime) [below= of dalc,xshift=-1.3cm,yshift=.1cm] {studytime};
                \node[state] (walc) [right= of dalc,xshift=-.2cm] {Walc};
                \node[state] (goout) [above= of walc,xshift=-0.5cm,yshift=-0.2cm] {goout};
                \node[state] (g1) [below= of walc,xshift=-0.2cm] {G1};
                \node[state] (g2) [right= of g1] {G2};
                \node[state] (g3) [above= of g2] {G3};

                \path (gender) edge (freetime);
                \path (gender) edge (dalc);
                \path (gender) edge (studytime);
                \path (gender) edge (g1);
                \path (freetime) edge (goout);
                \path (goout) edge (dalc);
                \path (dalc) edge (walc);
                \path (dalc) edge (g1);
                \path (studytime) edge (g1);
                \path (g1) edge (g2);
                \path (g2) edge (g3);
                \path (g1) edge (g3);
            \end{tikzpicture}
            \caption{Student Performance (Por)}\label{dag_por}
        \end{subfigure}
    \caption{Partial DAGs of the estimated causal model for educational datasets, showing only the sensitive attribute, its descendants, and the target variable. See Appendix \ref{append_graph} for full graphs. Each sub-graph is not used for implementing counterfactually fair models; only the remaining features are included.} \label{fig:causal models}
    \vspace{-1em}
\end{figure}

\subsection{Structural Causal Model of Educational Dataset}

Counterfactual fairness holds that intervening solely on the sensitive attribute A while keeping all other things equal, does not change the model's prediction distribution. To implement counterfactual fairness, a predefined Structural Causal Model(SCM) in Directed Acyclic Graph(DAG) form is necessary. Although the causal model of the Law School data exists\cite{kusner2017counterfactual}, there are no known causal models for the remaining datasets. 

To construct the SCM of OULAD and the Student Performance dataset, we use a causal discovery algorithm, Linear Non-Gaussian Acyclic Model (LiNGAM) \cite{shimizu2006linear}. The algorithm estimates a causal structure of the observational data of continuous values under linear-non-Gaussian assumption. From the estimated causal model, we filtered DAG weights that are under the 0.1 threshold.

Among constructed SCM, we present features that are in causal relationships with the sensitive attribute that directly or indirectly affects the target variable in Fig. \ref{fig:causal models}. Further analysis of causal relationships between sensitive features is discussed in Section \ref{section5}.

\subsection{Counterfactual Fairness Evaluation Metrics}

We use the Wasserstein Distance(WD) and Maximum Mean Discrepancy(MMD) metric for evaluating the difference between prediction distributions for sensitive attributes. Wasserstein distance and MMD are common metrics for evaluating counterfactual fairness\cite{duong2024achieving,ma2023learning}. Lower WD and MMD values suggest greater fairness, indicating smaller differences between the outcome distributions.

Although there exist other measures for evaluating counterfactual fairness such as Total Effect\cite{zuo2023counterfactually} and Counterfactual Confusion Matrix\cite{pinto2024matrix}, we limit our evaluation of counterfactual fairness to the above metrics. We construct unaware and counterfactual models without direct access to the sensitive attribute, evaluating fairness with mentioned metrics would not be feasible. We visually examine prediction distributions through Kernel Density Estimation(KDE) plots across our baseline and counterfactually fair models.

\subsubsection{Educational Domain Specific Fairness Metric}

We additionally analyze the counterfactual approach with pre-existing fairness metrics tailored for the education domain. We choose Absolute Between-ROC Area(ABROCA)\cite{gardner2019evaluating} and Model Absolute Density Distance(MADD)\cite{verger2023your} for the analysis. ABROCA quantifies the absolute difference between two ROC curves. It measures the overall performance divergence of a classifier between sensitive attributes, focusing on the magnitude of the gap regardless of which group performs better at each threshold. MADD constructs KDE plots of prediction probabilities and calculates the area between two curves of the sensitive attribute. While the ABROCA metric represents how similar the numbers of errors across groups are, the MADD metric captures the severity of discrimination across groups, allowing for diverse perspectives on the analysis of model behaviors on fairness. Although both metrics are designed for group fairness, we include those in our work because they are specifically proposed under the context of the educational domain. 

\subsection{Experiment Details}

For the experiment, we considered the Level 1 concept of counterfactual fairness defined in Kusner et al.\cite{kusner2017counterfactual}. At Level 1, the predictor is built exclusively using observed variables that are not causally influenced by the sensitive attributes. While a causal ordering of these features is necessary, no assumptions are made about the structure of unobserved latent variables. This requires causal ordering of features but no further assumptions of unobserved variables. For the Law School dataset, Level 2 is used.

We selected two baselines for the experiment, (a) Unfair model and (b) Unaware model. An unfair model directly includes sensitive attributes to train the model. The unaware model implements `Fairness Through Unawareness', a fairness notion where an algorithm is considered fair when protected attributes are not used in the decision-making process\cite{cornacchia2023auditing}. We compare two baselines with the FairLearning algorithm introduced in Kusner et al.\cite{kusner2017counterfactual}.

We evaluate the counterfactual fairness of machine learning models on both regression and classification models. We selected the four most utilized machine learning models in the algorithmic fairness literature\cite{hort2024bias}. We choose Linear Regression(LR; Logistic Regression for classification), Multilayer Perceptron(MLP), Random Forest(RF), and XGBoost(XGB)\cite{chen2016xgboost}. For KDE plot visualizations, we used a linear regression model for regression and MLP for classification.

\section{Result}
\label{section4}

In the result section of our study, we present an analysis of counterfactual fairness on educational datasets. Since the Law School dataset is well studied in the counterfactual fairness literature, we only provide this experiment as a baseline.

\subsection{Visual Analysis}

We use KDE plots to visualize outcome distributions across subgroups, providing a better understanding of counterfactual fairness with summary statistics.

\begin{minipage}{\linewidth}
\begin{figure}[H]
     \centering
     \begin{subfigure}[b]{0.325\textwidth}
         \centering
         \includegraphics[width=\textwidth]{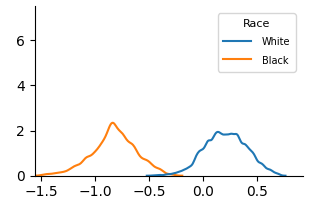}
         \caption{Unfair}
         \label{fig:law_unfair}
     \end{subfigure}
     \hfill
     \begin{subfigure}[b]{0.325\textwidth}
         \centering
         \includegraphics[width=\textwidth]{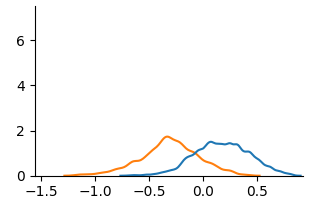}
         \caption{Unaware}
         \label{fig:law_unaware}
     \end{subfigure}
     \hfill
     \begin{subfigure}[b]{0.325\textwidth}
         \centering
         \includegraphics[width=\textwidth]{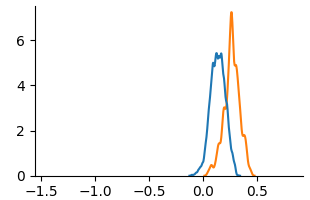}
         \caption{Counterfactual}
         \label{fig:law_cf}
     \end{subfigure}
    \caption{KDE plots on Law School.}\label{fig:kde_law}
\end{figure}

\vspace{-2em}

\begin{figure}[H]
     \centering
     \begin{subfigure}[b]{0.325\textwidth}
         \centering
         \includegraphics[width=\textwidth]{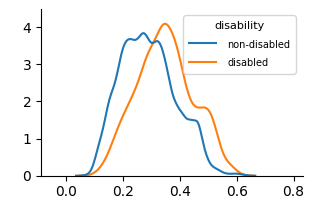}
         \caption{Unfair}
         \label{fig:oulad_unfair}
     \end{subfigure}
     \hfill
     \begin{subfigure}[b]{0.325\textwidth}
         \centering
         \includegraphics[width=\textwidth]{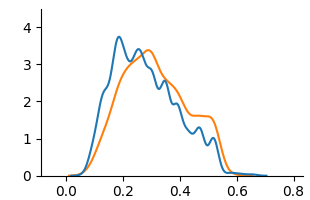}
         \caption{Unaware}
         \label{fig:oulad_unaware}
     \end{subfigure}
     \hfill
     \begin{subfigure}[b]{0.325\textwidth}
         \centering
         \includegraphics[width=\textwidth]{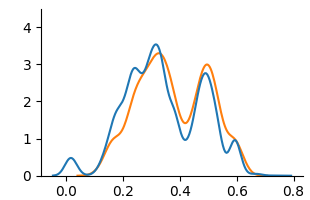}
         \caption{Counterfactual}
         \label{fig:oulad_cf}
     \end{subfigure}
    \caption{KDE plots on OULAD.}\label{fig:kde_oulad}
\end{figure}

\vspace{-2em}

\begin{figure}[H]
     \centering
     \begin{subfigure}[b]{0.325\textwidth}
         \centering
         \includegraphics[width=\textwidth]{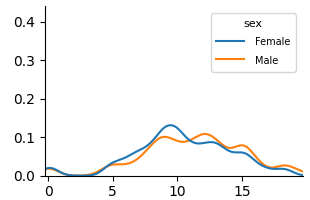}
         \caption{Unfair}
         \label{fig:mat_unfair}
     \end{subfigure}
     \hfill
     \begin{subfigure}[b]{0.325\textwidth}
         \centering
         \includegraphics[width=\textwidth]{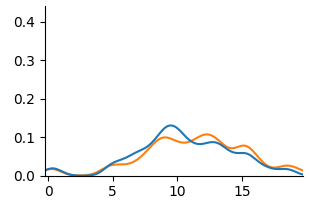}
         \caption{Unaware}
         \label{fig:mat_unaware}
     \end{subfigure}
     \hfill
     \begin{subfigure}[b]{0.325\textwidth}
         \centering
         \includegraphics[width=\textwidth]{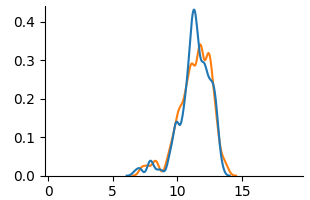}
         \caption{Counterfactual}
         \label{fig:mat_cf}
     \end{subfigure}
    \caption{KDE plots on Student Performance(Mathematics).}\label{fig:kde_mat}
\end{figure}

\vspace{-2em}

\begin{figure}[H]
     \centering
     \begin{subfigure}[b]{0.325\textwidth}
         \centering
         \includegraphics[width=\textwidth]{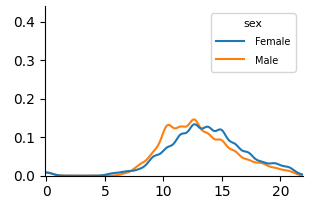}
         \caption{Unfair}
         \label{fig:por_unfair}
     \end{subfigure}
     \hfill
     \begin{subfigure}[b]{0.325\textwidth}
         \centering
         \includegraphics[width=\textwidth]{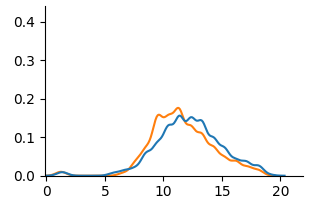}
         \caption{Unaware}
         \label{fig:por_unaware}
     \end{subfigure}
     \hfill
     \begin{subfigure}[b]{0.325\textwidth}
         \centering
         \includegraphics[width=\textwidth]{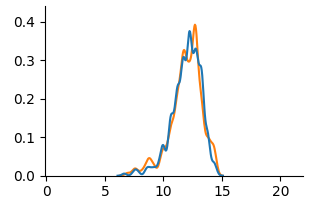}
         \caption{Counterfactual}
         \label{fig:por_cf}
     \end{subfigure}
    \caption{KDE plots on Student Performance(Portuguese).}\label{fig:kde_por}
\end{figure}
\end{minipage}

Fig. \ref{fig:kde_law} and Fig. \ref{fig:kde_oulad} present KDE plots for Law School and OULAD datasets. For Law School data, we can see that the unfair and unaware model produces predictions with large disparities, as previously known from the counterfactual literature. For OULAD data, unfair and unaware models' prediction probabilities do not overlap, giving slightly higher prediction probabilities for disabled students. For both datasets, the prediction distribution of the counterfactual model is closer than the unfair and unaware model. 

Fig. \ref{fig:kde_mat} and Fig. \ref{fig:kde_por} show KDE plots for Student Performance in Mathematics and Portuguese. The differences in model predictions are relatively small for all models compared to previous datasets, although disparities exist. In Mathematics, unfair and unaware models underestimate scores for female students (below 10) and overestimate for males. The opposite is true for Portuguese, where female students are more frequently assigned scores above 10. Counterfactual models on both data demonstrate an overlap of two distributions, although male students were predicted to be in the middle score range more frequently than female students in Mathematics.

\subsection{Measure of Counterfactual Fairness}

\vspace{-2.5em}

\begin{table}[h]
    \caption{Evaluation of fairness notions on benchmark datasets.}
    \label{fair_eval}
    \centering
    \begin{tabularx}{0.85\textwidth}{c|c|>{\centering\arraybackslash}X|>{\centering\arraybackslash}X|>{\centering\arraybackslash}X}
        \hline
        Data & Metric    & Unfair    & Unaware   & Counterfactual\\
        \hline
        \multirow{2}{*}{Law} & WD  & 1.0340 & 0.4685 & \textbf{0.1290} \\
         & MMD & 0.8658 & 0.4140 & \textbf{0.1277} \\
        \hline
        \multirow{2}{*}{OULAD} & WD  & 0.0722 & 0.0342 & \textbf{0.0337} \\ 
        & MMD & 0.0708 & 0.0324 & \textbf{0.0317} \\
        \hline
        \multirow{2}{*}{Math} & WD  & 0.7251 & 0.7358 & \textbf{0.1161} \\
        & MMD & 0.3396 & 0.1917 & \textbf{0.0538} \\
        \hline
        \multirow{2}{*}{Por} & WD  & 0.7526 & 0.6339 & \textbf{0.1047} \\
        & MMD & 0.4322 & 0.2839 & \textbf{0.1205} \\
        \hline
    \end{tabularx}
\end{table}

\vspace{-1em}

We present the evaluation of counterfactual fairness in Table \ref{fair_eval}. In all cases, the counterfactually fair model achieves the lowest WD and MMD. For Law School and Student Performance(Mat and Por) data, the distance between two distributions of sensitive attribute subgroups significantly reduced, comparing the counterfactual model to the unfair and unaware model. Despite the limited visual evidence of reduced distributional differences in the Student Performance KDE plots, WD and MMD provided quantifiable measures of this reduction. For OULAD data, the reduction in distribution difference between the unaware and counterfactual model is minimal, suggesting a weak causal link of \textmyfont{disability} to student's \textmyfont{final result}. Consistent for all datasets, WD and MMD decrease as the sensitive attribute and its causal relationships are removed.

Fairness levels vary across datasets. Law school data shows the highest initial unfairness while OULAD data shows relatively low unfairness even for the unfair model. Both Student Performance dataset shows significant unfairness, particularly for WD. WD and MMD rankings of fairness methods generally agree, with large differences in one corresponding to large differences in the other, suggesting robustness to the distance metric choice.

\vspace{-1.5em}
 
\begin{table}[h]
    \caption{Evaluation of education-specific fairness on OULAD dataset.}
    \label{clf_eval}
    \centering
    \begin{tabularx}{0.85\textwidth}{c|c|>{\centering\arraybackslash}X|>{\centering\arraybackslash}X|>{\centering\arraybackslash}X}
        \hline
        Data & Metric    & Unfair    & Unaware   & Counterfactual\\
        \hline
        \multirow{2}{*}{OULAD} & ABROCA & 0.1019 & 0.0219 & \textbf{0.0181} \\ 
        & MADD & 0.5868 & 0.3194 & \textbf{0.2763} \\
        \hline
    \end{tabularx}
\end{table}

\vspace{-1em}

Given the classification nature of the OULAD dataset, ABROCA and MADD metric results are presented in Table \ref{clf_eval}. Because ABROCA and MADD assess group fairness disparity across all classification thresholds, they are not directly comparable to counterfactual fairness, an individual-level fairness notion. However, the unfair model was highly biased, as evidenced by its ABROCA (0.1019, max 0.5) and MADD (0.5868, max 2) scores. While the unaware model showed improvement, the counterfactual model achieved the best fairness results. This indicates that the counterfactual approach is effective in reducing disparities in the number of errors and model behaviors across groups.

\subsection{Performance of Machine Learning Models}

\vspace{-2em}

\begin{table}[h]
    \caption{Prediction performance of machine learning models on fairness notions.} \label{reg_perf}
    \centering
    \begin{tabularx}{\textwidth}{c|
                                 c|
                                 >{\centering\arraybackslash}X
                                 >{\centering\arraybackslash}X
                                 >{\centering\arraybackslash}X
                                 >{\centering\arraybackslash}X|
                                 >{\centering\arraybackslash}X
                                 >{\centering\arraybackslash}X
                                 >{\centering\arraybackslash}X
                                 >{\centering\arraybackslash}X|
                                 >{\centering\arraybackslash}X
                                 >{\centering\arraybackslash}X
                                 >{\centering\arraybackslash}X
                                 >{\centering\arraybackslash}X}
         \hline
         \multirow{2}{*}{Data} &  \multirow{2}{*}{Metric} & \multicolumn{4}{c|}{Unfair} & \multicolumn{4}{c|}{Unaware} & \multicolumn{4}{c}{Counterfactual} \\
           &   & LR & MLP & RF & XGB & LR & MLP & RF & XGB & LR & MLP & RF & XGB \\ 
         \hline
         \multirow{2}{*}{Law} & MSE & 0.72 & 0.73 & 0.50 & 0.52 & 0.75 & 0.75 & 0.59 & 0.59 & 0.82 & 0.83 & 0.57 & 0.57 \\
           & RMSE & 0.85 & 0.86 & 0.71 & 0.72 & 0.86 & 0.86 & 0.77 & 0.77 & 0.90 & 0.91 & 0.75 & 0.76 \\
         \hline
         \multirow{2}{*}{OULAD} & Acc & 0.69 & 0.70 & 0.72 & 0.71 & 0.69 & 0.69 & 0.71 & 0.71 & 0.68 & 0.68 & 0.70 & 0.69  \\
           & AUROC & 0.65 & 0.68 & 0.72 & 0.71 & 0.65 & 0.67 & 0.71 & 0.70 & 0.62 & 0.63 & 0.65 & 0.64  \\ 
         \hline
         \multirow{2}{*}{Mat} & MSE & 4.13 & 5.33 & 2.82 & 4.71 & 4.06 & 5.30 & 2.88 & 4.04 & 17.43 & 17.50 & 17.08 & 17.76 \\
           & RMSE & 2.03 & 2.31 & 1.68 & 2.17 & 2.01 & 2.30 & 1.70 & 2.01 & 4.17 & 4.18 & 4.13 & 4.21 \\
         \hline
         \multirow{2}{*}{Por} & MSE & 1.43 & 1.74 & 2.19 & 1.60 & 1.41 & 1.42 & 2.17 & 1.73 & 7.96 & 8.61 & 7.90 & 8.52 \\
           & RMSE & 1.20 & 1.32 & 1.48 & 1.27 & 1.19 & 1.19 & 1.47 & 1.31 & 2.82 & 2.93 & 2.81 & 2.92  \\
         \hline
    \end{tabularx}
\end{table}

\vspace{-1em}

We show model performance results in Table \ref{reg_perf}. Across models, tree-based ensembles (RF and XGB) generally outperformed LR and MLP in regression. LR and MLP showed variable performance, with strong results on the Law School dataset but poor performance on others. All models performed well on the Law School dataset; however, the Student Performance datasets (Mathematics and Portuguese) were more challenging, possibly due to non-linear relationships.

The impact of fairness approaches varies across datasets. Although the unfair model frequently has the highest performance, the classification performance of OULAD remains similar across all fairness approaches. For Law School and Student Performance data, the counterfactual model leads to the worst performance, which aligns with existing literature on the accuracy-fairness trade-off. Student Performance in Mathematics shows a massive increase in MSE and RMSE for all models, suggesting that achieving counterfactual fairness with performance is challenging on this dataset. 

\section{Discussion}
\label{section5}

\subsubsection{RQ 1. What causal relationships do sensitive attributes have in educational data?}

\begin{figure}[!htbp]
    \centering
    \includegraphics[width=\textwidth]{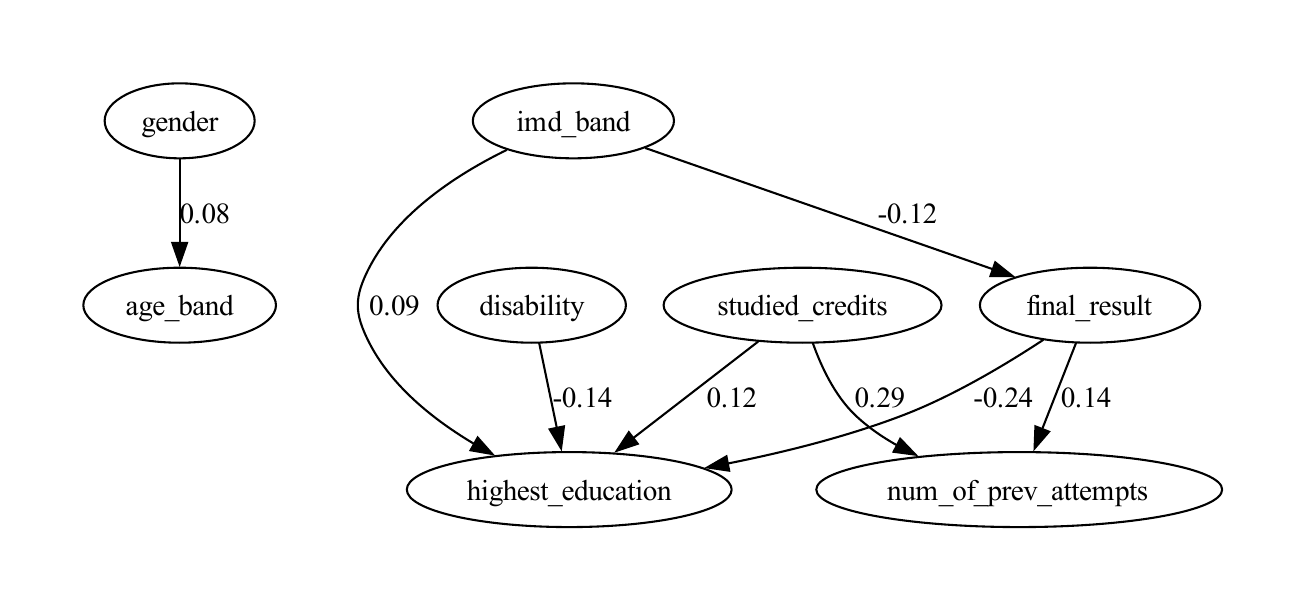}
    \caption{Estimated SCM for OULAD dataset. Sensitive attribute is \textmyfont{disability}. For fitting counterfactual model, we excluded \textmyfont{disability} and \textmyfont{highest\_education} features.}
    \label{fig:scm_oulad}
\end{figure}

Analysis of the OULAD causal graph (Fig. \ref{dag_oulad} and Fig. \ref{fig:scm_oulad}) reveals that \textmyfont{disability} has a direct causal effect on \textmyfont{highest education} (-0.14 weight). This implies that having \textmyfont{disability} makes attaining \textmyfont{higher education} more difficult. There is no common cause between \textmyfont{disability} and \textmyfont{final result}, implying having a disability does not directly affect student outcome. Attribute \textmyfont{gender} causally affects \textmyfont{age}; however, with a 0.1 edge weight threshold, two attributes are disconnected from the DAG. This reinforces previous research\cite{hasan2022understanding} which revealed no causal relationship between \textmyfont{gender} and \textmyfont{final result}.

The causal model of Student Performance is presented in Fig. \ref{dag_mat} and Fig. \ref{dag_por}. The estimated causal model shows potential gender-based influences in study habits, social behaviors, and alcohol consumption to academic performance. Foremost, \textmyfont{gender} have an indirect causal relationship on \textmyfont{G3}. For both datasets, \textmyfont{gender} directly influences \textmyfont{studytime}, and \textmyfont{studytime} directly influences \textmyfont{G1}. For Mathematics, \textmyfont{gender} directly impacts \textmyfont{studytime}, \textmyfont{freetime}, \textmyfont{goout} and \textmyfont{Dalc}, but not \textmyfont{goout} for Portuguese. Differences in \textmyfont{goout} and alcohol consumption(\textmyfont{Dalc} and \textmyfont{Walc}) show that the factors influencing student performance differ between Math and Portuguese, demonstrating the importance of considering subject-specific causal models in education. 

\vspace{-0.5em}

\subsubsection{RQ 2. Does counterfactual fairness in educational data lead to identical outcomes for students regardless of their demographic group membership in individual-level?}

From our experiment result, we have demonstrated that removing causal links between sensitive attributes and the target through counterfactuals achieves a similar prediction distribution of machine learning models in sensitive feature subgroups. This suggests that the counterfactual approach is effective at mitigating unfairness as measured by these metrics, across all datasets. 

The fairness result supports the insufficiency of the `fairness through unawareness' notion in educational datasets. In KDE plots from Fig. \ref{fig:kde_law} to Fig. \ref{fig:kde_por}, (a) Unfair are often very similar to (b) Unaware. In fairness evaluation in Table \ref{fair_eval} and Table \ref{clf_eval}, the Unaware approach generally performs better than the Unfair baseline, but it's significantly worse than the Counterfactual approach. This suggests that proxies often exist within the remaining features and simply removing the sensitive attribute is not a reliable way to achieve fairness.

\vspace{-0.5em}

\subsubsection{RQ 3. Does counterfactually fair machine learning models result in a performance trade-off in educational data?}

The performance result in Table \ref{reg_perf} demonstrates trade-off exists between achieving high predictive accuracy and satisfying counterfactual fairness, especially for Student Performance data. Although the definition of counterfactual fairness is agnostic to how good an algorithm is\cite{kusner2017counterfactual}, this phenomenon is known from the previous literature\cite{zhou2024counterfactual} that an trade-off between fairness and accuracy exists dominated by the sensitive attribute influencing the target variable. 

The severe performance drop in the Student Performance dataset suggests high dependence on sensitive attribute \textmyfont{gender} on student performance, especially for Mathematics subject. We can infer that machine learning models heavily rely on the information related to the sensitive attribute \textmyfont{gender} for prediction. Removal of the sensitive attribute and its causal influence can drastically reduce performance in this case. 

Similar performance across all fairness approaches in the OULAD dataset implies that sensitive attribute \textmyfont{disability} might not be a significant feature for predicting student outcomes. Further, the naive exclusion of sensitive attributes has minimal impact on the performance of machine learning models, reconfirming the ineffectiveness of the Unaware approach in both fairness and performance.

Overall, we find the nature of the sensitive attribute and its causal links to other features differs across educational datasets, influencing the variability in the effectiveness of the counterfactual fairness approach. Some sensitive attributes might be more challenging to address than others in terms of counterfactual fairness. 

\vspace{-0.5em}

\subsubsection{Limitations and Future Work}

Our work is limited to implementing the early approach of counterfactual fairness introduced in Kusner et al.\cite{kusner2017counterfactual}, which only includes non-descendants of sensitive attributes in the decision-making process and utilizing the Level 1 causal model. Also, we only report on counterfactual fairness and performance trade-offs. Thus, future research will focus on developing our Level 1 causal model into a Level 2 model. This will involve postulating unobserved latent variables based on expert domain knowledge and assessing the impact of increasingly strong causal assumptions. Concurrently, we will develop algorithms to reduce the trade-off between counterfactual fairness and performance in educational datasets.

\vspace{-0.5em}

\section{Conclusion}
\label{section6}

In this paper, we evaluated the counterfactual fairness of machine learning models on real-world educational datasets and provided a comprehensive analysis of counterfactual fairness in the education context. This work contributes to exploring causal mechanisms in educational datasets and their impact on achieving counterfactual fairness. Considering counterfactual fairness as well as group and individual fairness could provide different viewpoints in evaluating the fairness of algorithmic decisions in education.

\begin{credits}
\subsubsection{\ackname}
We acknowledge the valuable input from Sunwoo Kim, whose comments helped in conducting the experiments. This work was supported by the National Research Foundation(NRF), Korea, under project BK21 FOUR (grant number T2023936). 

\subsubsection{\discintname}
The authors have no competing interests to declare that are
relevant to the content of this article.
\end{credits}


\bibliographystyle{splncs04}
\bibliography{mybibliography}

\newpage

\begin{subappendices}
\renewcommand{\thesection}{\Alph{section}}%

\section{Feature Description of Student Performance Dataset} \label{append_data}

\begin{table}[H]
    \caption{Feature descriptions of Student Performance dataset\cite{cortez2008using}.}
    \label{dataset:student performance}
    \centering
    \begin{tabularx}{\textwidth}{s s w}
        \hline
        Feature & Type & Description\\
        \hline
            school & binary & the student's school (Gabriel Pereira/Mousinho da
Silveira) \\
            gender & binary & The student's gender \\
            age & numerical & The student's age \\
            address & binary & The student's residence (urban/rural)\\
            famsize & binary & The student's family size\\
            Pstatus & binary & The parent's cohabitation status\\
            Medu & numerical & Mother's education \\
            Fedu & numerical & Father's education \\
            Mjob & categorical & Mother's job \\
            Fjob & categorical & Father's job \\
            reason & categorical & The reason to choose this school \\
            guardian & categorical & The student's guardian (mother/father/other) \\
            traveltime & numerical & The travel time from home to school \\
            studytime & numerical & The weekly study time \\
            failures & numerical & The number of past class failures \\
            schoolsup & binary & Is there an extra educational support? \\
            famsup & binary & Is there any family educational support? \\
            paid & binary & Is there an extra paid classes within the course subject? \\
            activities & binary & Are there extra-curricular activities? \\
            nursery & binary & Did the student attend a nursery school? \\
            higher & binary & Does the student want to take a higher education? \\
            internet & binary & Does the student have an Internet access at home? \\
            romantic & binary & Does the student have a romantic relationship? \\
            famrel & numerical & The quality of family relationships \\
            freetime & numerical & Free time after school  \\
            goout & numerical & How often does the student go out with friends? \\
            Dalc & numerical & The workday alcohol consumption \\
            Walc & numerical & The weekend alcohol consumption \\
            health & numerical & The current health status \\
            absences & numerical & The number of school absences \\
            G1 & numerical & The first period grade \\
            G2 & numerical & The second period grade \\
            G3 & numerical & The final grade \\
        \hline
    \end{tabularx}
\end{table}

\newpage
\section{Complete SCMs of Datasets}\label{append_graph}

Construction of causal structural model(SCM) is crucial for implementing counterfactual fairness. Thus, we provide a estimated SCM inferred from each dataset through LiNGAM algorithm\cite{shimizu2006linear}. We filtered out edges with absolute weights lower than 0.1. These causal models are used for sampling counterfactual instances. For fitting a counterfactually fair model, we excluded direct and indirect descendants of the sensitive feature for each dataset.

\begin{figure}[!htbp]
    \centering
    \includegraphics[width=\textheight,angle=90,origin=c]{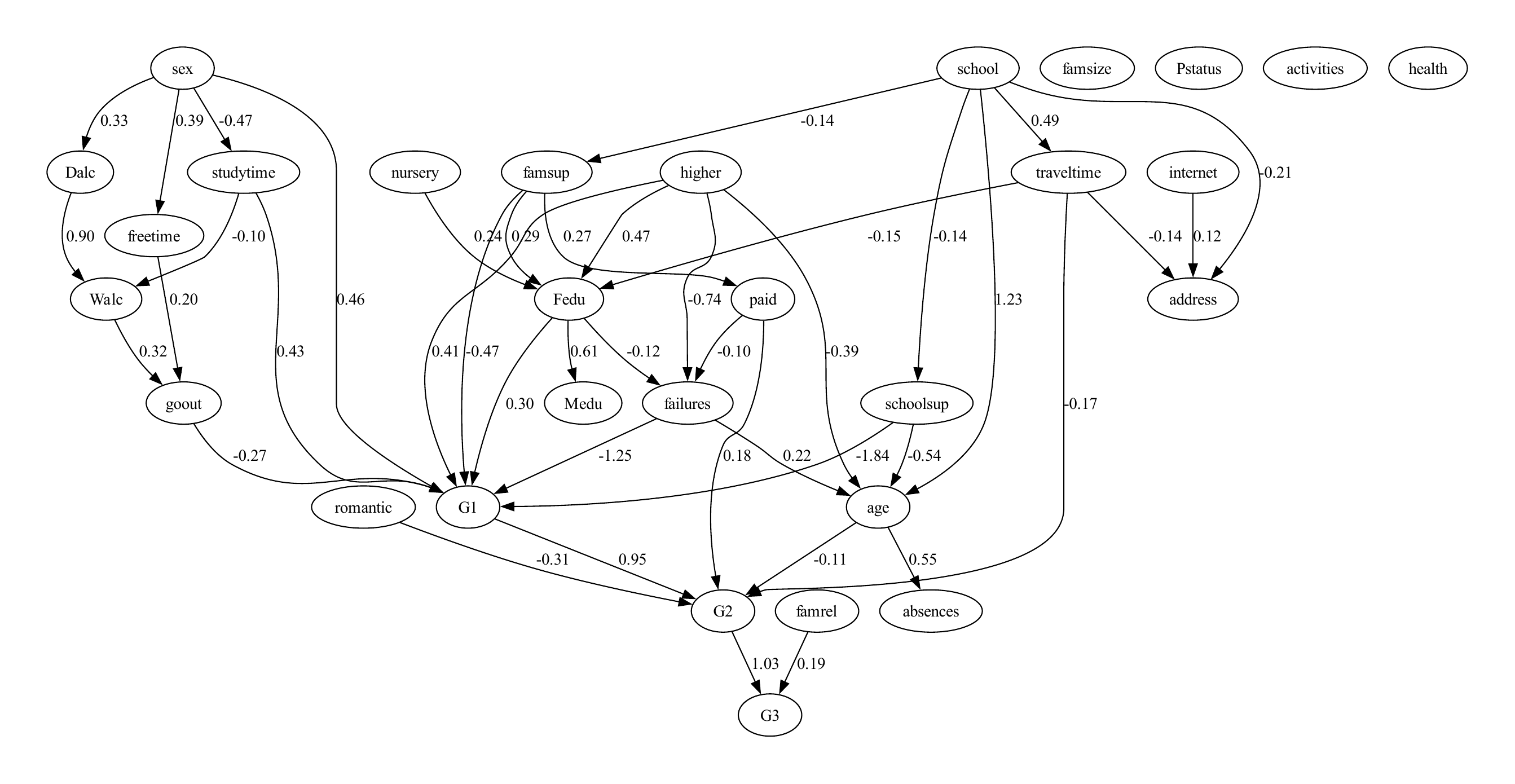}
    \caption{Estimated SCM for Student Performance(Mathematics) dataset. Sensitive attribute is \textmyfont{gender}. For fitting counterfactual model, we excluded \textmyfont{gender}, \textmyfont{freetime}, \textmyfont{goout}, \textmyfont{Dalc}, \textmyfont{Walc}, \textmyfont{famsup}, \textmyfont{paid}, \textmyfont{G1}, \textmyfont{G2}, \textmyfont{absences} and \textmyfont{studytime}. Features that does not have edge connected to the rest of the graph are also excluded.}
    \label{fig:scm_student_mat}
\end{figure}

\begin{figure}[!htbp]
    \centering
    \includegraphics[width=\textheight,keepaspectratio,angle=90,origin=c]{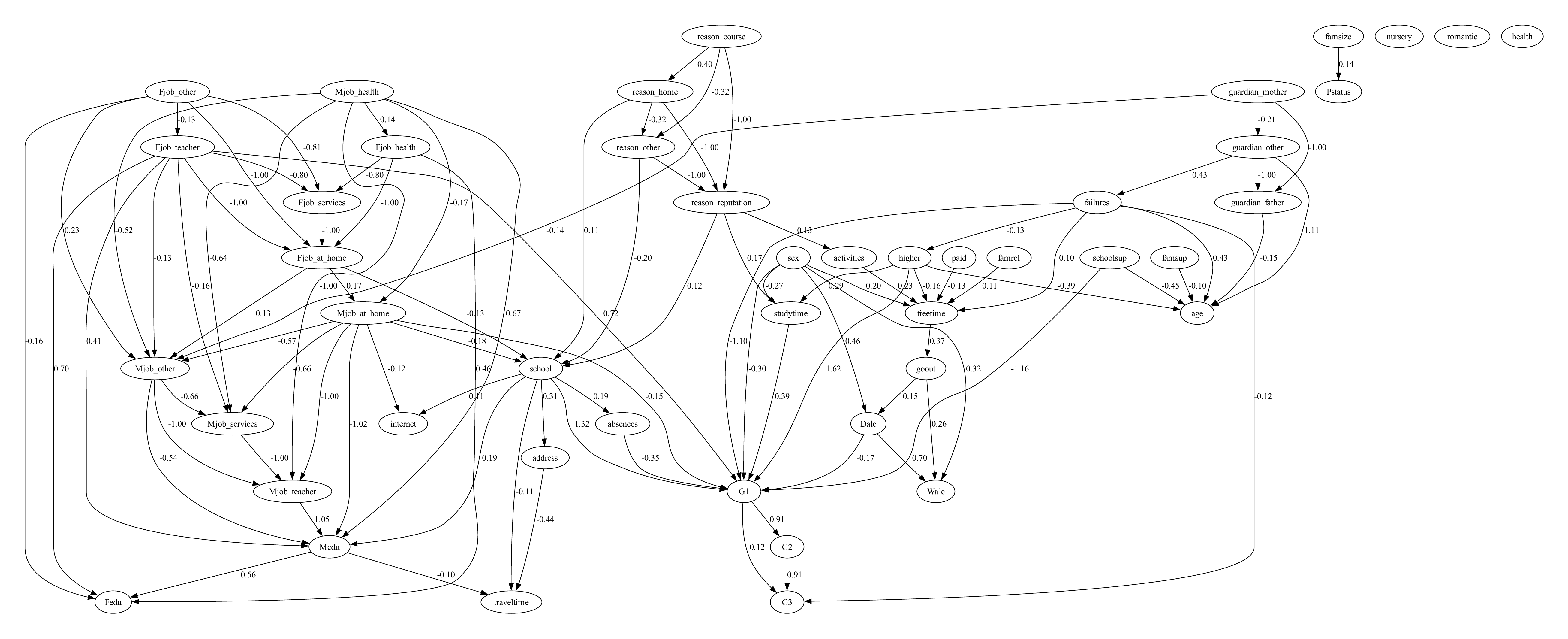}
    \caption{Estimated SCM for Student Performance(Portuguese) dataset. Sensitive attribute is \textmyfont{gender}. For fitting counterfactual model, we excluded \textmyfont{gender}, \textmyfont{freetime}, \textmyfont{goout}, \textmyfont{Dalc}, \textmyfont{Walc}, \textmyfont{G1}, \textmyfont{G2}, \textmyfont{absences} and \textmyfont{studytime}. Features that does not have edge connected to the rest of the graph are also excluded.}
    \label{fig:scm_student_por}
\end{figure}

\end{subappendices}

\end{document}